# Simulation and performance study of ceramic THGEM[*]

YAN Jia-Qing(颜嘉庆)[1,2;1)] XIE Yu-Guang(谢宇广)[2,3;2)] HU Tao[2,3](胡涛)

LU Jun-Guang[2,3](吕军光) ZHOU Li[2,3](周莉) QU Guo-Pu[1](屈国普)

CAI Xiao[2,3](蔡啸) NIU Shun-Li[2,3](牛顺利) CHEN Hai-Tao[2](陈海涛)

[1]University of South China, Hengyang 421001, China

[2]State Key Laboratory of Particle Detection and Electronics, Beijing 100049, China

[3]Institute of High Energy Physics, CAS, Beijing 100049, China

**Abstract:** THGEMs based on a ceramic substrate have been successfully developed for neutron and single photon detection. The influences on thermal neutron scattering and internal radioactivity of both ceramic and FR-4 substrates were studied and compared. The ceramic THGEMs are homemade, of 200 um hole diameter, 600 um pitch, 200 um thickness, 80 um rim, and 50mm × 50mm sensitive area. FR-4 THGEMs with the same geometry were used as a reference. The gas gain, energy resolution and gain stability were measured in different gas mixtures using 5.9 keV X-rays. The maximum gain of a single layer ceramic THGEM reaches $6\times10^4$ and $1.5\times10^4$ at $Ne + CH_4$ = 95:5 and $Ar + i\text{-}C_4H_{10}$ = 97:3, respectively. The energy resolution is better than 24%. Good gain stability was obtained during a more than 100 hour continuous test in $Ar+CO_2$ = 80:20. By using a $^{239}Pu$ source, the alpha deposited energy spectrum and gain curve of the ceramic THGEM were measured.

## 1. Introduction

Thick Gaseous Electron Multipliers (THGEMs), a geometrical expansion of standard GEMs, are robust, cheap, and easy to manufacture and have a high gain [1-3]. THGEMs are fabricated by mechanical drilling based on standard printed circuit board (PCB) technology. Over the past decade, as one type of new micro pattern gaseous detectors (MPGDs), there has been continual study and development of THGEM detectors. Besides the well-known applications, such as Ring Imaging Cherenkov (RICH) detectors [4-5] and charged particle detectors [6-7], THGEMs are also very useful for cesium iodide (CsI)-based single photon detection [8,9] and neutron detection [10], because of their high gain and sub-millimeter spatial resolution. These applications require the THGEM to have low internal radioactivity and low neutron scattering.

Generally, THGEMs are made of FR-4 substrate, a popular type of PCB substrate. Ceramic substrate is different from FR-4, and is expected to have lower neutron scattering, less out-gassing and lower internal radioactivity. These characteristics are suitable for single photon and neutron detection, such as in gaseous photo multipliers (GPMs) and thermal neutron beam monitors, as mentioned above. Both FR-4 and ceramic substrates are composite materials. The different characteristics are due to the different components. Pure ceramic is mainly composed of clay, quartz and feldspar, and the ceramic substrate is made from pure ceramic mixed with 10% glass

Supported by National Natural Science Foundation of China (11205173) and the State Key Laboratory of Particle Detection and Electronics (H9294206TD)

1) E-mail: yanjq@ihep.ac.cn

2) E-mail: ygxie@ihep.ac.cn (corresponding author)

fiber, as shown in Table1. The highest content of elements in the ceramic is Oxygen, about 48.5%. The neutron scattering cross section of Oxygen is only 3.8 b [11]. Hydrogen, on the other hand, for which the neutron scattering cross section is 47.7 b, has a concentration of almost zero in the ceramic substrate.

Table1. Composition of the ceramic substrate

| Ceramic substrate | | | | | | | | |
|---|---|---|---|---|---|---|---|---|
| Composition | pure ceramic(90%) | | | | | | | glass fiber |
| Element | O | Si | Al | Fe | Mg | Ca | Na | K |
| mass ratio | 43.5% | 30.3% | 8% | 2% | 0.1% | 0.2% | 2.4% | 4.5% | 10% |

In this paper, the influences on thermal neutron scattering of both FR-4 and ceramic substrates were simulated and compared using Geant4. The internal radioactivity of both substrates was also measured by a low-background high-purity germanium (HPGe) spectrometer. The performance of the newly developed ceramic THGEMs was studied in detail.

## 2. Geant4 simulation

For neutron detection, the spatial resolution required is at millimeter level, such as 1~3mm, which is the advantage given by using THGEMs. However, the neutron detection efficiency is the crucial problem which needs to be improved for THGEM-based neutron detectors. The solution, naturally, is to adopt a cascade structure of neutron convertors plus THGEMs. In this case, the thickness of the THGEM and the neutron scattering and absorption of the substrate must be considered, especially for cold and thermal neutrons. Fig.1 is an example for the cascade structure. Under each THGEM surface, a layer of boron-10 is plated, with an optimized thickness of 2 um. The boron is coated on the bottom but not the top surface of the THGEM, as this may be more effective for the gas ionization and electron incidence to the holes due to the electric field.

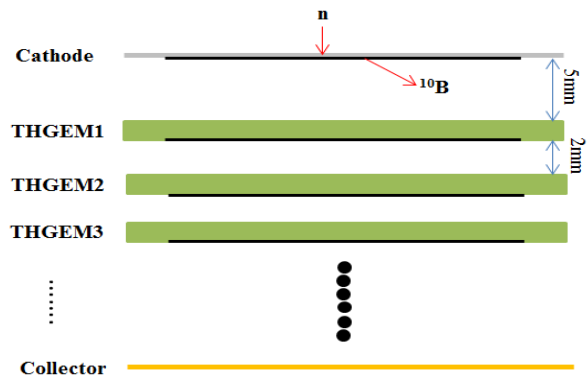

Fig.1. Schematic view of the THGEM neutron detection

### 2.1 Scattering simulation

By using Geant4 and its physics list for low energy neutron interactions (QGSP_ BERT_HP), the scattering effects can be simulated for both FR-4 and ceramic substrates. In the simulation, the neutron energy was set as 0.025 eV, the typical value of thermal neutrons, and the substrates had no holes. Fig. 2 shows the scattering radius distributions of thermal neutrons through two layers of ceramic or FR-4 substrates with a thickness of 0.2 mm for each layer and a 2 mm gas gap. The inducing or collecting plane is placed 2 mm behind the last layer of substrate, to locate the positions of scattered neutrons. The total number of incident neutrons is the same for both substrates. The distributions after normalization by area, i.e. the total number of incident neutrons, indicate that the most probable values (MPVs) of scattering radii for both substrates are almost the same, about 2 mm, which will affect the position uncertainty greatly. The scattering radii in FR-4 mostly centralize at the peak, while those in ceramic spread more to the tail. However, if the distributions are normalized by the total number of incident neutrons but not the number of scattered neutrons, it can be seen that more neutrons, about 3 times, are scattered by the FR-4 than by the ceramic.

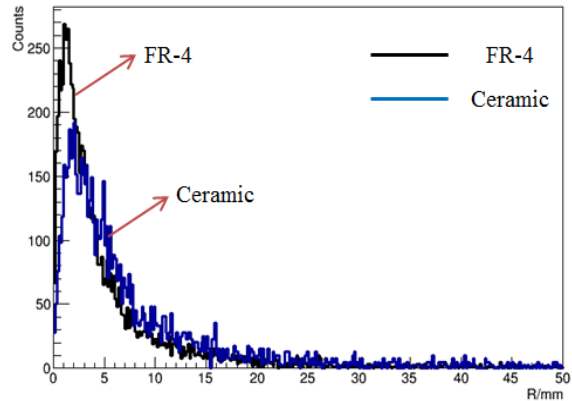

Fig.2. The scattering radius distributions of the thermal neutrons

through two layers of ceramic and FR-4 substrates(normalized by area)

The neutron loss includes both scattered and absorbed parts. Fig. 3 shows the loss and absorption ratios of thermal neutrons as a function of substrate layer. As can be seen, both neutron loss and absorption ratios are proportional to the layer for both ceramic and FR-4 substrates. The absorption ratio in ceramic is little lower than that in FR-4, while the loss ratio in FR-4 is much higher than that in ceramic, about two to three times. This means that the neutron scattering is more severe in FR-4 than in ceramic substrate. In addition, the impact on neutron loss is dominated by scattering rather than absorption for both substrates. Fig. 4 shows the thermal neutron scattering ratio as a function of substrate layer by different cuts of scattering radius, i.e. $R_{cut}$ of 0.2, 1 and 1.5 mm. It is obvious in Fig. 4 that the scattering ratios are proportional to the layer for both ceramic and FR-4. For the same $R_{cut}$, the scattering ratio in FR-4 is several times higher than that in ceramic. For the ceramic substrate, the scattering ratios are almost the same for $R_{cut}$ = 1.5 mm, 1.0 mm and 0.2 mm.

The simulation results indicate that scattering probability, radius and absorption probability due to the substrate should be considered when the THGEMs are adopted for thermal neutron detection. Both the loss ratio and the scattering ratio of the ceramic substrate are much lower than those of FR-4.

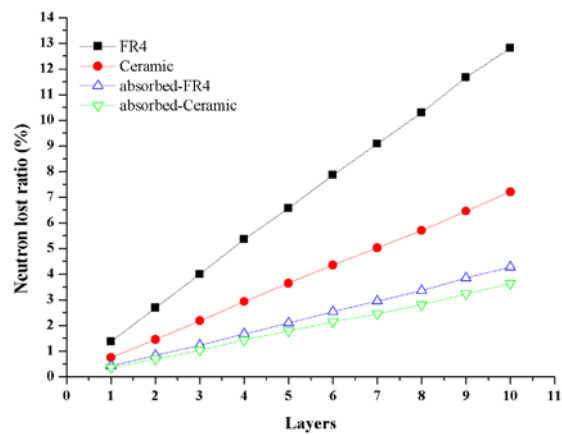

Fig. 3. The thermal neutron lost ratios as a function of

ceramic and FR-4 substrate layers

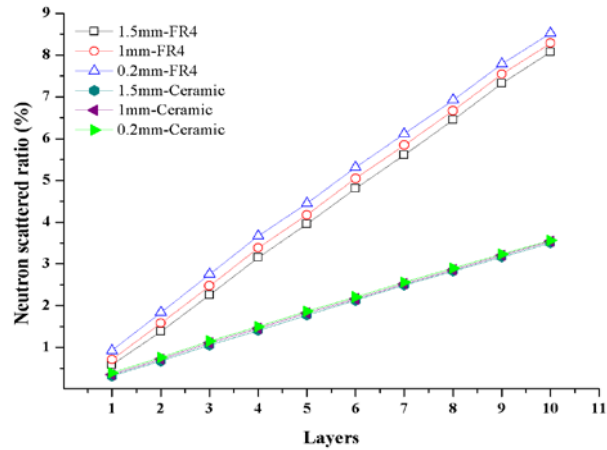

Fig. 4.The thermal neutron scattering ratio as a function of ceramic

and FR-4 substrate layers ( scattering radius >=$R_{cut}$, $R_{cut}$ =0.2, 1 or 1.5 mm)

## 2.2 Detection efficiency and position resolution simulation

Since alpha ionization produces about one thousand times more primary electrons than X-rays, the detection efficiency for neutrons is dominated by the neutron-alpha conversion efficiency. By using Geant4, the relationship between the thick of boron-10 and the alpha conversion efficiency was simulated. The thermal neutrons enter the center of a boron-10 layer vertically, and the collecting plane is placed 0.3 mm behind the boron-10 layer to locate the positions of α particles. The variation in alpha conversion efficiency was studied by changing the thickness of boron-10. As Fig. 5 shows, in the beginning, the alpha conversion efficiency increases exponentially with the increase inboron-10 thickness. The alpha conversion efficiency reaches its maximum, about 3%, when the thickness is about 2.5 um. This process is dominated by the alpha conversion itself. When the layer of boron-10 is made thicker and thicker, more and more alphas are absorbed by the boron-10 layer and can not be emitted into the gas, so the total conversion efficiency decreases slowly instead of increasing continually.

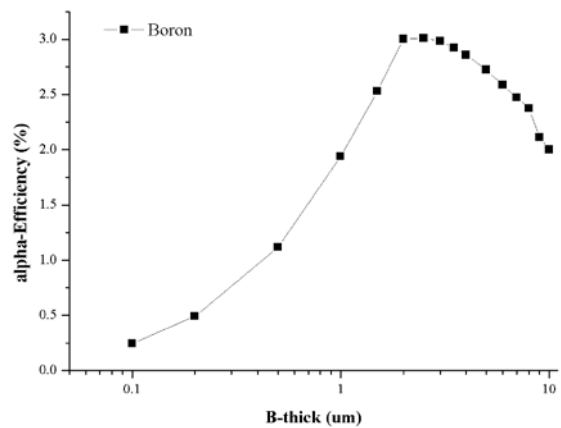

Fig.5. Alpha conversion efficiency as a function of the thickness of boron-10

We then simulated the detection efficiency and position resolution for the cascade THGEM neutron detector. The FR-4 or ceramic substrates without holes were built with the same geometry, and the thickness of boron-10 was set as 2 um and coated on the bottom surfaces of each THGEM. The collecting plane was placed 2 mm behind the last layer substrate to locating the positions of α particles. Since the alpha detection efficiency is almost 100%, the detector was placed in vacuum rather than gas, in order to avoid the influence of gas absorption on emitting α. Here, the position resolution of the THGEM neutron detector is reflected by the sigma of the scattering radius distributions of thermal neutrons, as shown in Fig. 6. There is no doubt that the conversion efficiency will be improved by more conversion layers; the simulation results confirm this and show that it is not simply linear and independent of the substrates. The cost of more conversion layers, however, is worse position resolution. For the same number of substrate layers, the position resolution using the ceramic is about 1mm worse than that using FR-4, due to the more high-Z elements in ceramic. For five layers, the conversion efficiency is more than 12%, and the position resolution is 2 and 3 mm for FR-4 and ceramic substrates, respectively.

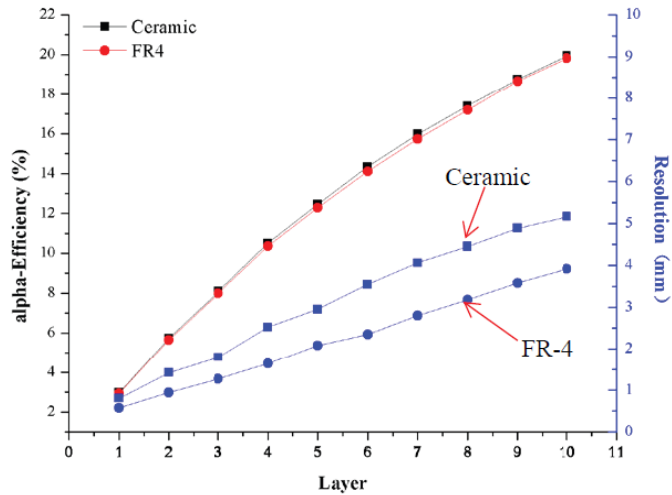

Fig.6. Alpha conversion efficiency and position resolution as a function

of number of layers of boron-10 coated THGEMs

## 2.3 Gamma rejection ratio simulation

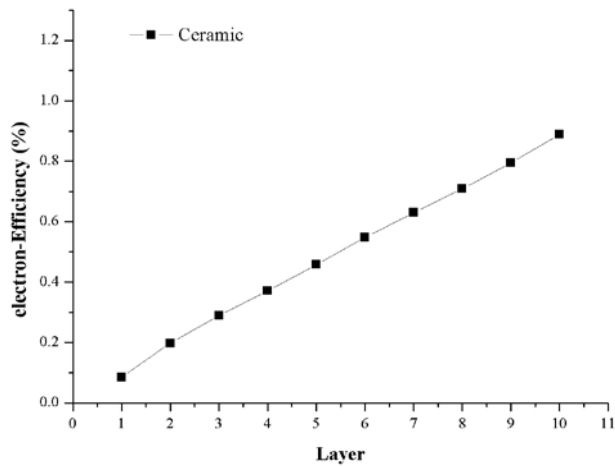

Fig.7. Gamma (1 MeV) induced electron production ratio as a function

of number of layers of ceramic substrate

The gamma background is the main concern for neutron detectors. For gas detectors, the probability of gamma (>1 MeV) induced gas ionization is very low, and the major part of the background comes from the interaction between gamma rays and the substrates, which will produce electrons with several hundred keV energy and so become the background. Therefore, the gamma rejection ratio depends mostly on the electron production ratio by gamma rays. In the simulation, 1 MeV gammas were used to study the electron production ratio in the cascade structure. The ceramic substrates were 2 mm thick and plated with 2 um boron-10 on the bottom surfaces. Fig. 7 gives the result of electron production ratio as a function of the number of layers of ceramic substrates. The electron production ratio is proportional to the number of substrate layers; this is another cost of increasing the neutron conversion efficiency by using a cascade structure. It can be seen in Fig. 7 that about 1 MeV gammas produce electrons and contaminate into the signals. Then, for 5 substrate layers, the gamma rejection ratio is about 4.6‰.

## 3. Radioactivity measurement

In the low background laboratory at the Institute of High Energy Physics (IHEP), Beijing, the internal radioactivity of ceramic and FR-4 substrates was analyzed using a high-purity germanium (HPGe) detector. The HPGe detector has 40% relative efficiency and crystal volume of 170 cm$^3$. The background level of this system is about 1 cps/100 cm$^3$Ge (50 keV~2.8 MeV), after combining the shielding and veto. Both ceramic and FR-4 substrate samples were cut to the same shape, without copper cladding. By measuring the gamma deposited energy spectrum and comparing with the standard deposited spectrum, the radioactive elements in the materials, such as $^{238}$U, $^{232}$Th and $^{40}$K, can be found. The specific radioactivity of the radioactive element can be calculated by the formula: R = Counts/ηtM, where R is the specific radioactivity, Counts is the counts at full energy peak, η is the detection efficiency of the HPGe detector, t is the testing time and M is the mass of the tested sample. Table 2 shows the specific radioactivity of the ceramic and FR-4 substrates. From Table 2, it can be seen that the specific radioactivity of the ceramic substrate is lower than that of FR-4. The specific radioactivity from all $^{232}$Th, $^{238}$U, and $^{40}$K of FR-4 substrates is about three times higher than those of ceramic substrates. Obviously, the ceramic THGEMs are a better choice for low background detection, with the specific radioactivity of $^{40}$K being 9.31±0.64 Bq/Kg.

Table 2: Specific radioactivity of ceramic and FR-4 substrates

|  | $^{232}$Th | | $^{238}$U | | $^{40}$K | |
| --- | --- | --- | --- | --- | --- | --- |
| Specific radioactivity | (Bq/Kg) | σ | (Bq/Kg) | σ | (Bq/Kg) | σ |
| Ceramic | 9.01 | 0.47 | 7.43 | 1.66 | 9.31 | 0.64 |
| FR-4 | 27.22 | 0.84 | 19.60 | 4.32 | 21.92 | 1.35 |

## 4. Performance of the ceramic THGEM

In this work, the ceramic THGEMs are homemade, with thickness of 0.2 mm, hole diameter of 0.2 mm, pitch of 0.6 mm and rim of 80 um. The ceramic substrate was developed specially for superior high frequency applications in the PCB industry. In our case, the attractive features are low out gassing, low neutron scattering, and low internal radioactivity. Both homemade and Rogers4000 series ceramic substrates were used for making THGEMs and tested. In order to compare the THGEM performances effectively, both ceramic and FR-4 THGEMs were produced with the same geometric parameters and with a 5 × 5 cm$^2$ active area. Fig. 8 shows the THGEM samples of ceramic and FR-4 substrates.

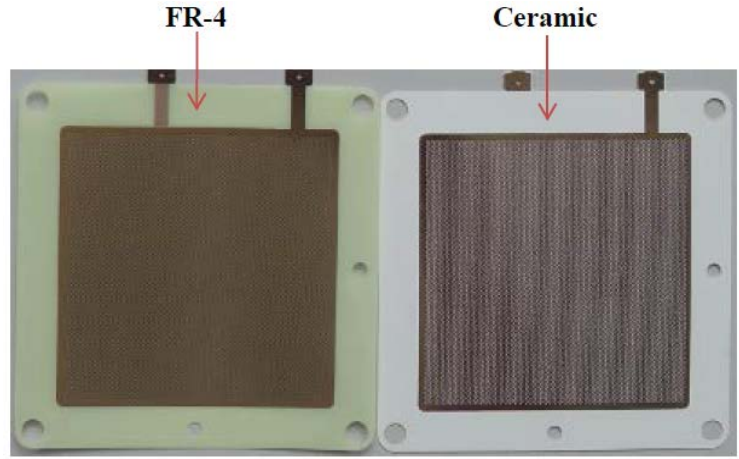

Fig. 8.THGEM samples of ceramic and FR4 substrates with the same geometry

### 4.1 Experimental setup

For basic performance tests, the test configuration is shown in Fig. 9. The test chamber consists of cathode, anode and a single layer THGEM. The measurements were carried out at room temperature in gas flowing mode. A $^{55}$Fe X ray source (10 mCi) was placed on the top of the chamber and collimated by a ф2 hole. The X-rays entered the upper drift region perpendicularly, where primary ionization was produced by 5.76 keV photo electrons, which were excited by the interaction of $^{55}$Fe 5.9 keV X-rays with Ar atoms. The ionization electrons were amplified in avalanche mode in the THGEM holes, then entered the lower inductive region, where they were finally collected by the anode. The high voltages (HVs) were supplied by a CAEN Mod.N470 mini-HV power. The signals were readout by an Ortec 142AH charge-sensitive preamplifier, 450 primary amplifier and TRUMPPCI-8K Multi-Channel Analyzer (MCA). The system was calibrated with an Ortec 415 pulse generator and a standard capacitor (2pC/mV) [12].

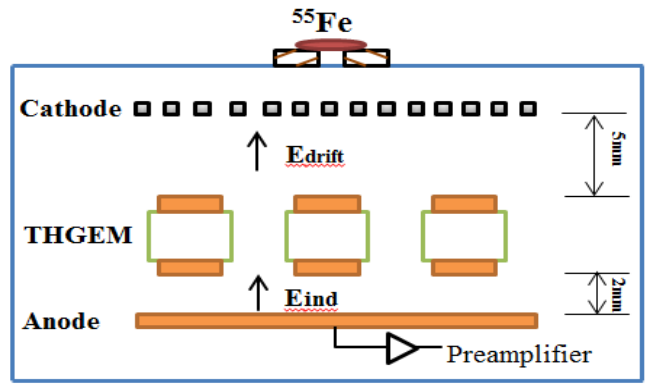

Fig. 9.Schematic view of the THGEM test chamber

### 4.2 Effective gain

Gain is the most important performance index for MPGDs. The HV working range and gain range can be given by the gain curve, which is the basis to set up the operating HV and design the back-end readout electronics. The gain curves were measured and compared in several popular gas mixtures: $Ne + CH_4 = 95:5$, $Ar + i\text{-}C_4H_{10} = 97:3$ and $Ar + CO_2 = 80:20$. The induction field Ei and the drift field Ed were set as 4.0 kV/cm and 1.0 kV/cm, respectively. The gas gaps are 5 mm and 2 mm for drift and induction field, respectively. The effective gain was measured with the voltage increments of 20 V or 10 V until spark discharge occurred. Fig. 10 shows the gain curves of single-layer THGEMs with both ceramic and FR-4 substrates. As Fig. 10 shows, the ceramic and FR-4 THGEMs have similar HV working range, which is around or longer than 100V. The maximum effective gains of FR-4 THGEM are higher than those of ceramic THGEM in almost all gas mixtures, which indicates the advantages of FR-4 substrate, which is the most popular and mature type of THGEM substrate. As a new substrate, the ceramic THGEM presents a promising gain performance, which still has space for improvement. For neutron detection, since the detected particles are alphas, the effective gain doesn't need to be too high, around 1000 is enough. From this point of view, the ceramic THGEM is excellent. The maximum gain of ceramic THGEM can reach $5\times10^4$ in Ne + $CH_4$ = 95:5, $1\times10^4$ in Ar + i-$C_4H_{10}$ = 97:3, and $5\times10^3$ in Ar + $CO_2$ = 80:20, respectively.

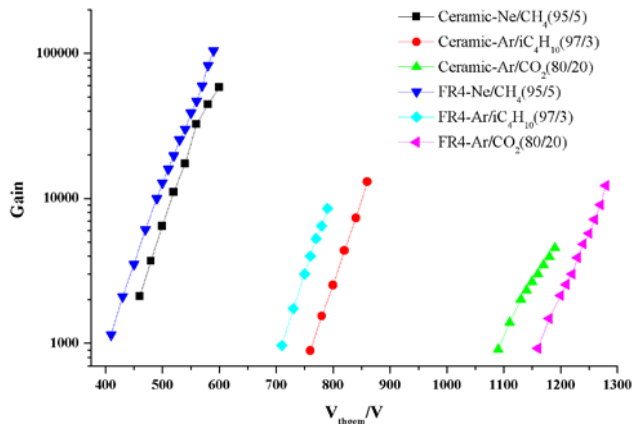

Fig.10. Effective gain of single THGEMs made of ceramic and

FR-4 substrate in different gas mixtures.

### 4.3 Energy resolution

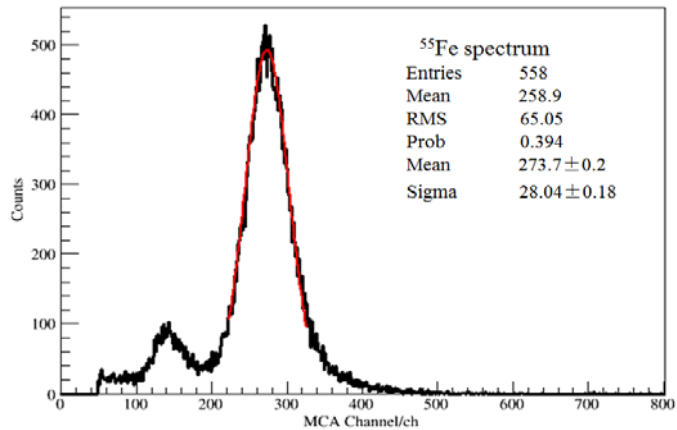

Fig.11. A typical $^{55}$Fe energy spectrum, single ceramic THGEM inAr/i$C_4H_{10}$ (97/3) gas mixture,

Ed = 1 kV/cm, Ei = 4 kV/cm and V$_{THGEM}$ = 800 V. The energy resolution is 24.04%.

With gas detectors, it can be difficult to obtain very high energy resolution. However, the energy resolution can reflect the quality of micro cells in MPGDs, as resolution depends on machining precision and uniformity. High energy resolution is therefore also an important goal for MPGD development. The most convenient way to get the energy resolution for MPGDs is to measure the energy spectrum of 5.9 keV $^{55}$Fe X-rays in the argon-based gas mixtures. The 5.9 keV $^{55}$Fe X-rays produce a full energy peak at 5.76 keV, and at the same time, an escape peak at 2.9 keV, so in the spectrum, two peaks should be seen, with the position of the full energy peak being just twice that of the escape peak. The energy resolution can be obtained by fitting the full energy peak. Fig. 11 shows a typical energy spectrum for a single layer ceramic THGEM in gas mixture of Ar + i-$C_4H_{10}$ = 97:3. By fitting the full energy peak with a Gaussian function, the sigma is obtained, then the full width at the half maximum (FWHM) can be calculated by the formula FWHM = 2.35 × σ. The energy resolution is then FWHM/Peak × 100%. For a single layer ceramic THGEM, the energy resolution measured currently is 24.04% at gain $3×10^3$, which is a normal level, with the possibility of further improvement.

**4.4 Gain stability**

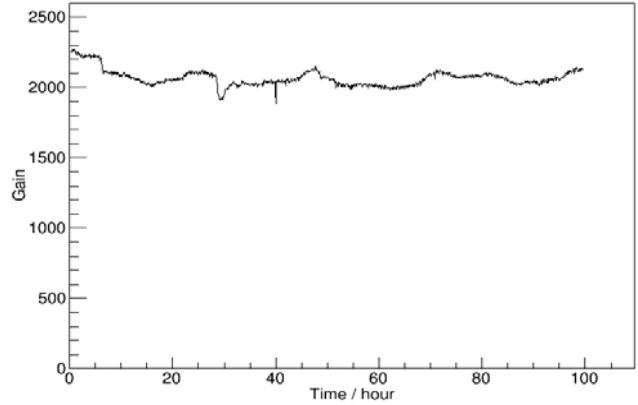

Fig. 12. The gain stability of single layer ceramic THGEM in Ar/CO$_2$ (80/20) for more than 100 hours,

Ed = 1 kV/cm, Ei = 4 kV/cm and V$_{THGEM}$ = 1150V, and the gain is about 2100.

The gain stability of THGEM is a key problem for its practical applications, especially in the cases where high gain and high rate are required. In order to certify the gain stability of the homemade ceramic THGEMs, we monitored the gain fluctuation on a relatively longtime scale. Fig. 12 shows the gain stability test result for a single ceramic THGEM using $^{55}$Fe X-rays in Ar + CO$_2$ = 80:20 for more than 100 hours. The voltage applied at the THGEM, i.e. V$_{THGEM}$, is 1150 V, and the $^{55}$Fe energy spectrum was acquired by MCA every 5 minutes. As Fig. 12 shows, at the beginning, the gain was about 2300, and then decreased a little. After 30 hours, the gain was almost steady at 2050 and continued so to the end of the test. The gain fluctuation was small and the variation was about 200/2050 = 9.76%, which is within 10%. Therefore, for neutron detection, in which case the gain required is less than 1000, the gain stability of the ceramic THGEM is assumed. For single photon detection, in which case the gain required is at the $10^5$ level, the gain stability should be tested by practical applications.

**4.5 α response**

The robustness and sub-millimeter spatial resolution properties of the THGEM are well suited to neutron detection, no matter whether fast or thermal neutrons [13]. Since neutrons cannot be detected directly by gas ionization, THGEM based neutron detectors must be combined with some kind of neutron convertors such as $^3$He, $^{10}$B, or $^6$Li. The ultra-high price of $^3$He gas in recent years has stimulated people to find an alternate and effective method for neutron detection. Therefore the THGEM and convertor cascade structure was proposed. For thermal neutrons, $^{10}$B (n, α) is a good convertor candidate. It reacts with neutrons as shown in the following Eq.1, Eq.2 and Eq.3 [14]:

$$^{10}B + n \longrightarrow {}^7Li + \alpha + 2.792 \text{MeV} \quad (1)$$

$$^{10}B + n \longrightarrow {}^7Li^* + \alpha + 2.310 \text{MeV} \quad (2)$$

$$^7Li^* \longrightarrow {}^7Li + \gamma \quad (3)$$

The boron-10 converts neutrons into 1.78 MeV and 1.47 MeV α particles, so for the THGEM itself, the particles which need to be detected are alphas, which will produce many more primary ionization electrons than X-rays. It also means that high gain is not necessary for neutron detection. However, the detection efficiency and neutron/gamma separation power are crucial for THGEM based neutron detectors, since there is a high gamma background accompanying the neutron beam. Fig. 13(b) shows the gain curves of single ceramic THGEM tested in Ar + $CO_2$ = 80:20 with 5/2 mm drift/induction gas gap by $^{239}$Pu α particles and $^{55}$Fe X-rays. It can be seen clearly that the applied voltage is nearly 300V lower for 5.5MeV α particles than for 5.9 keV $^{55}$Fe X-rays to reach the same ADC channel. So, by adjusting the gain and operation voltage, the neutron signals can be separated effectively from X-rays. As we know that the photo electron effect is dominant in energy <=10 MeV, and the cross section decreases with the energy increase, so the high energy (MeV level) gamma background should be much lower than for X-rays. Fig. 13 (a) is the 5.5 MeV α spectrum at a gain of about 28, and the voltage applied is 850 V. The α deposited energy distribution is very clear, with a Landau shape.

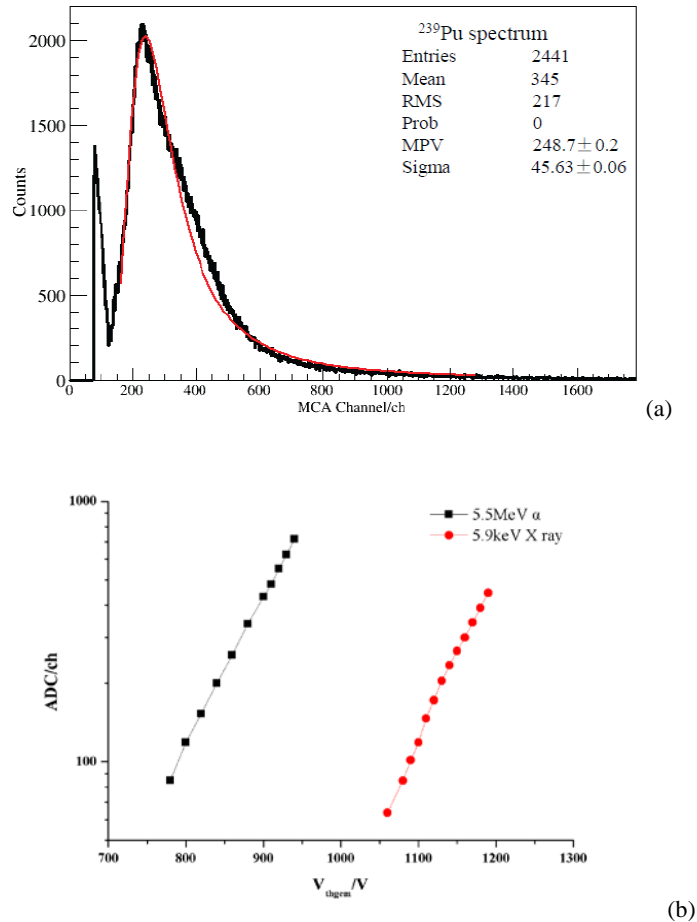

Fig. 13.Test results forα particles in Ar + $CO_2$ = 80:20. (a) The α deposited energy spectrum recorded by single ceramic THGEM. (b) The high voltage operation range of THGEM for α and X-rays

## 5. Conclusions

The homemade ceramic THGEMs show good performance for single photon and thermal neutron detections. The internal radioactivity of ceramic substrate is only one third of that of FR-4. The neutron loss ratio with ceramic substrate is only half to one third of that with FR-4, and the loss ratio is dominated by the scattering ratio, rather than the absorption ratio. With five ceramic substrate layers, the detection efficiency of THGEM neutron detection is more than 12%, the position resolution is 2.95 mm, and the gamma rejection ratio is 4.6‰. The maximum gas gain of a single layer ceramic THGEM reached about $5\times10^4$ in Ne + $CH_4$ = 95:5, and $5\times10^3$ in Ar + $CO_2$ = 80:20. The energy resolution is currently 24.04% in Ar + i-$C_4H_{10}$ = 97:3 at gain of about $3\times10^3$. The gain can keep good stability for more than 100 hours in Ar + $CO_2$ = 80:20, which is the favorite gas mixture for neutron detection. Using a $^{239}$Pu α source, the ceramic THGEM showed good response for alpha particles, and so should have a good response for neutrons after conversion. Due to the 300 V difference in high voltage operation range between α particles and X-rays, the neutron signals are expected to be easily distinguished from the gamma background.

A ceramic THGEM with thinner substrate and bigger sensitive area is still under development. For practical applications, the ideal thickness is 0.1mm, and the sensitive area should be larger than 200×200 mm$^2$.

## Acknowledgement

This work is supported in part by the National Natural Science Foundation of China (11205173) and the State Key Laboratory of Particle Detection and Electronics (H9294206TD). We acknowledge the support and cooperation of Huizhou King Brother Circuit Technology Co., Ltd (KBC).